\documentclass[showpacs,prl,superscriptaddress,nofootinbib, twocolumn]{revtex4}
\usepackage{graphicx}
\usepackage{color}
\usepackage{amssymb,amsfonts,amsmath}
\usepackage{natbib}
\usepackage{epstopdf}
\usepackage{subfigure}
\begin{document}

\title{Prestrain-induced contraction in 1D random elastic chains}

\author{Ihusan Adam}  
\affiliation{Department of Information Engineering, University of Florence, Florence, Italy}
\affiliation{Department of Physics and Astronomy, and CSDC, University of Florence, Sesto Fiorentino, Italy}

\author{Franco Bagnoli} 
\affiliation{Department of Physics and Astronomy, and CSDC, University of Florence, Sesto Fiorentino, Italy}
\affiliation{INFN, Florence Section, Sesto Fiorentino, Italy}

\author{Duccio Fanelli}  
\affiliation{Department of Physics and Astronomy, and CSDC, University of Florence, Sesto Fiorentino, Italy}
\affiliation{INFN, Florence Section, Sesto Fiorentino, Italy}

\author{L. Mahadevan}
\affiliation{School of Engineering and Applied Sciences, Department of Physics, and Department of Organismic and Evolutionary Biology, Harvard University, Cambridge, MA 02138, USA}

\author{Paolo Paoletti}
\affiliation{School of Engineering, University of Liverpool, L69 3GH Liverpool, UK}

\begin{abstract}
Prestrained elastic networks arise in a number of biological and technological systems ranging from the cytoskeleton of cells to tensegrity structures. To understand the response of such a network as a function of the prestrain, we consider a minimal model in one dimension. We do this by considering a chain (1D network) of elastic springs upon which a random, zero mean, finite variance  prestrain is imposed. Numerical simulations and analytical predictions quantify the magnitude of the contraction as a function of the variance of the prestrain, and show that the chain always shrinks. To test these predictions, we vary the topology of the chain and consider more complex connectivity and show that our results are relatively robust to these changes.
\end{abstract}


\maketitle

\section{Introduction}

Mass-spring networks or Elastic Network Models (ENMs) have been studied extensively due to their potential to model many interesting real-world phenomena. Earlier applications of such networks have been in the simulation of deforming bodies such as clothes and other fabric-like materials for computer graphics \cite{Hutchinson1996AdaptiveSimulations,Howlett1998Mass-SpringPoints,Bayraktar2007PracticalCloth}.  Recently, soft deforming bodies of a biological nature have been studied using ENMs with applications ranging from the study of facial tissue \cite{Duysak2003EfficientSystems} to simulation of body organs in a surgical context \cite{Zerbato2007CalibrationSimulations}. Even smaller micro-scale biological phenomena, such as the membrane mechanics of cells, have benefited from the simplicity of ENMs \cite{Chen2014InvestigationModelling,Rausch2013OnMembranes}. This is in part due to the efficiency and low computational complexity of ENMs compared to other competing simulation frameworks such as finite elements methods \cite{Delhomme2007SimulationSystem}.

When it comes to the study of large molecular structures such as proteins, coarse-grained ENMs have been shown to be computationally more efficient than the more accurate atomic models which do not scale as well with the number of atoms or molecules to be simulated \cite{Atilgan2001AnisotropyModel}. In such studies, the nodes of the ENM act as single amino acid residues, while the links model the inter-residue potentials. Despite the obvious simplifications introduced in their formulation, ENMs have been shown to capture not only the folding and unfolding of conformations, but also fluctuations around these shapes \cite{Yang2009ProteinCooperativity,Echave2008EvolutionaryModel,Dietz2008}. The normal modes of vibration ENMs can even describe  the large-amplitude motions related to ATP binding and hydrolysis in various molecular machines and motors \cite{Togashi2007NonlinearMachines,Wieninger2011ATPAnalysis,Zheng2003AModel}. 

Inspired by protein-machines, there is a growing trend to design mechanical networks that perform specific prescribed tasks \cite{Flechsig2018DesignedMachinery}. Abstract networks are also at the basis of many coordination algorithms in robotics, such as consensus and formation control   \cite{CORTES2017CoordinatedSurvey}.

With potential interest ranging from material sciences to biophysical systems to robotics, in this paper we propose a novel method that uses a random, zero-mean prestrain to induce contractions in 1D elastic networks. Prestrain is an important concept in many diverse fields, from tissue and bio-film engineering \cite{Rausch2013OnMembranes,Duysak2003EfficientSystems} in the biological realm, to the design of elastomer-based artificial muscle actuators \cite{Brochu2013All-siliconeRobustness,Ha2007InterpenetratingStrain,Kim2017ANetwork,Ha2006InterpenetratingMuscles} in the realm of robotics. 

Starting from an intuitive explanation as to the nature of this contraction within the proposed framework, we develop an analytical theory to show how the strength of the noise affects the final contraction. We also explore how the topology of these contacting networks affect the final state by numerically testing Watts-Strogatz \cite{Watts1998CollectiveNetworks} networks and scale-free networks \cite{Caldarelli2007Scale-FreeTechnology} of, respectively, varying rewiring probability and degree exponent. It is found that the amount of contraction depends mainly on the magnitude of the noise, and the effect of the topology fades as the size of the network increases, i.e. in the thermodynamic limit.

\section{Problem statement}
 
We consider a network composed of point masses linked by linear springs with a spring constant $k=\frac{\gamma}{l_0}$, where $l_0$ is the rest length of the spring and $\gamma$ can be regarded as the mechanical stiffness of the springs (proportional to the Young's modulus of the material composing it). The positions, $x_i$, of $N$ masses are initialised at a regular spacing over a line segment of length $[-5N, 5N]$\footnote{The choice of $5$ is arbitrary, however it is important for the length of the system to scale linearly with $N$ for meaningful comparisons of networks of different sizes.}, so that they are equidistant. In the following, the connections among masses - corresponding to springs in the ENM - are generated according to the Watts-Strogatz method \cite{Watts1998CollectiveNetworks}, unless otherwise stated. By doing so, networks of increasing disorder or randomness can be created by starting from a regular lattice (of N nodes) with a specified number of links to each neighbour. Each link is then rewired randomly with a set rewiring probability $p$.

Once the positions of the masses have been initialised and the links between them have been assigned, we set the rest length $l_{0_{ij}}$ of the spring connecting masses $j$ and $i$, such that the system is at equilibrium, i.e. $l_{0_{ij}}=\left|x_j-x_i\right|$.

A prestrain is then applied to the system by changing the spring rest length as $l_{0_{ij}} \mapsto \hat{l}_{0_{ij}} =l_{0_{ij}}(1+\delta_{ij})$, and $\hat{l}_{0_{ij}}$ refers to the new rest length after perturbation. For each spring, the prestrain term $\delta_{ij}$ is drawn from a random uniform distribution over the range $\delta \in [-\delta_{max},\delta_{max}]$. The system is then allowed to relax to its new equilibrium. 

The dynamics of the system can be modelled by the set of ordinary differential equations

\begin{equation}
    \ddot{x}_i=\sum_j A_{ij}\gamma \left(\frac{1}{\hat{l}_{0_{ij}}} - \frac{1}{l_{ij}} \right) (x_j - x_i) - \eta\dot{x_i} - F_{r_{ij}} 
    \label{eq1}
\end{equation}
where $l_{ij}=|x_j - x_i|$ and $\mathbf{A}$ is the adjacency matrix of the network, i.e. $A_{ij}=A_{ji}=1$ if there exists a spring between masses $i$ and $j$ and $A_{ij}=A_{ji}=0$ otherwise. Dissipation has been modelled by adding a viscous term with coefficient $\eta$. The term $F_{r_{ij}}$ models simple rigid sphere interaction, preventing the masses from switching their relative ordering along the segment as the system relaxes towards its equilibrium state. For the simulation described below, this term is set to $F_{r_{ij}}=(x_j-x_i)^{-3}$. However, as the theoretical analysis will show below, the exact form of this term does not matter, as long as masses do not switch order during relaxation.

Given that links share the same stiffness $\gamma$, i.e. have a spring constant $k$ that depends on the rest length, a prestrained network tends to  contract when subject to the prestrain described above. This is because the zero mean perturbation of rest length $l_{0_{ij}} \mapsto \hat{l}_{0_{ij}} =l_{0_{ij}}(1+\delta_{ij})$ creates a perturbed distribution of spring constants

\begin{equation}
  \hat{k}_{ij}=\frac{\gamma}{l_{0_{ij}}(1+\delta_{ij})}
\end{equation}
that is biased. In other words, the uniform distribution of prestrain  $P(\delta)$ results in a distribution  $P(k)$ of spring constants that is skewed more towards stronger and shorter springs. Hence, springs with negative prestrain ($\delta_{ij}<0$) become stiffer than springs with positive prestrain and the overall network is expected to shrink.% when $\hat{l}_0<l_0$ and weaker springs when $\hat{l}_0>l_0$, i.e $P(k_{ij})=\frac{\gamma}{k_{ij}^2 l_{0_{ij}}}$.

\section{Expected Shrinkage}
An exact analytical solution for the steady-state solution of Eq.~\eqref{eq1} is difficult to obtain due to the presence of the nonlinear term $F_{r_{ij}}$. Therefore, here only the expected value of the steady-state position $x_i$ is studied. This allows the analysis of the mean outcome of the problem within given bounds.

To this end, let us note at first that the problem can be simplified by making use of the fact the nodes cannot swap positions during motion thanks to the term $F_{r_{ij}}$.
Let us then define $\dot{x}=v_i$ and rewrite the model in Eq.~\eqref{eq1} in matrix form as 
\begin{equation}
\dot{\overrightarrow{V}} = \gamma (\mathbf{L} \overrightarrow{X} - \overrightarrow{B}) - \eta\overrightarrow{V}
\label{eq3}
\end{equation}
where each element of the vector $\overrightarrow{B}$ is $B_i= \sum_j{A_{ij}}~ sign\{x_j - x_i\}$ and each element of the Laplacian operator matrix $\mathbf{L}$ is defined as 
\begin{equation}
    L_{ij}=A'_{ij} - k'\delta_{ij} 
\label{eq4}
\end{equation}
Here, the following quantities are defined to simplify the notation: $A'_{ij}=\frac{A_{ij}}{l_{0_{ij}}}$ and $k'_i = \sum_jA'_{ij}$. 

Note that the vector $\overrightarrow{B}$ is a constant as long as the masses keep their relative ordering, i.e. do not swap positions, during relaxation. Furthermore, the repulsion term $F_{ij}$ that models the rigid sphere collisions in Eq.~\eqref{eq1} has been omitted in Eq.~\eqref{eq3}. This was done on the assumption that this repulsive force is active only on a very short range, so that it does not significantly alter the final state of the system. In other words, we assumed that the effect of the rigid sphere collisions can be adequately captured by having a fixed vector $\overrightarrow{B}$ in the following analysis. This assumption will be validated \emph{a posteriori} by comparing the predictions from the analytical theory with numerical tests.

It is important to note that the system is in equilibrium  $\left(\overrightarrow{V}_0,\dot{\overrightarrow{V}}_0\right)=(0,0)$ before the prestrain is applied, therefore the initial positions $\overrightarrow{X_0}$ must satisfy 
\begin{equation}
L_0 \overrightarrow{X_0} = \overrightarrow{B}
\label{eq5}
\end{equation}

where $L_0$ is the initial, unperturbed Laplacian matrix from Eq.~\eqref{eq4} formed from the initial rest lengths $l_{0_{ij}}$.

The expected final equilibrium position  can be estimated by setting $\dot{\overrightarrow{V}}=\overrightarrow{V}=0$, so that the Eq.~\eqref{eq3} reads
\begin{equation}
L(\delta) \overrightarrow{X_f} = \overrightarrow{B}
\label{eq6}
\end{equation}
where $X_f$ denotes the final positions of the masses.

Upon perturbation, the only term that is directly affected by the prestrain $\delta$ is the Laplacian \textbf{L}. Hence, by taking the expected value of both sides of Eq.~\eqref{eq6}, one obtains 
\begin{equation}
\langle L(\delta) \rangle \langle \overrightarrow{X_f} \rangle = \overrightarrow{B} 
\label{eq7}
\end{equation}

By integrating each term over the noise distribution one can then estimate the average equilibrium positions of the $N$ masses. To this end, let us consider at first $\langle L_{ij} \rangle = \int_{-\delta_{max}}^{\delta_{max}} L_{ij}(\delta) P(\delta) d\delta$ with $P(\delta)=\frac{1}{2\delta_{max}}$, thus obtaining
\begin{equation}
\langle L_{ij} \rangle =\psi L_{0_{ij}}
\label{eq8}
\end{equation}
where the shrinkage factor $\psi$ is defined as
\begin{equation}
\psi = \frac{1}{2\delta_{max}}ln\left(\frac{1+\delta_{max}}{1-\delta_{max}}\right)
\label{eq9}
\end{equation}

Combining Eq.~\eqref{eq7} with Eq.~\eqref{eq8} then yields
\begin{equation}
 L_0 \psi \langle \overrightarrow{X_f} \rangle = \overrightarrow{B} 
\label{eq10}
\end{equation}

The expected relative shift of each mass can thus be found by comparing Eq. \eqref{eq10} and Eq. \eqref{eq5} as

\begin{equation}
\frac{1}{\psi} = \frac{\langle x_{f_i} \rangle}{x_{0_i}}
\label{eq11}
\end{equation}
where $x_{0_i}$ and $x_{f_i}$ are the $i$-th elements of the initial position vector $\overrightarrow{X_0}$ and expected final position vector $\langle \overrightarrow{X_f} \rangle$, respectively. 

\begin{figure}[htb!]
    \centering
   \includegraphics[scale=0.425]{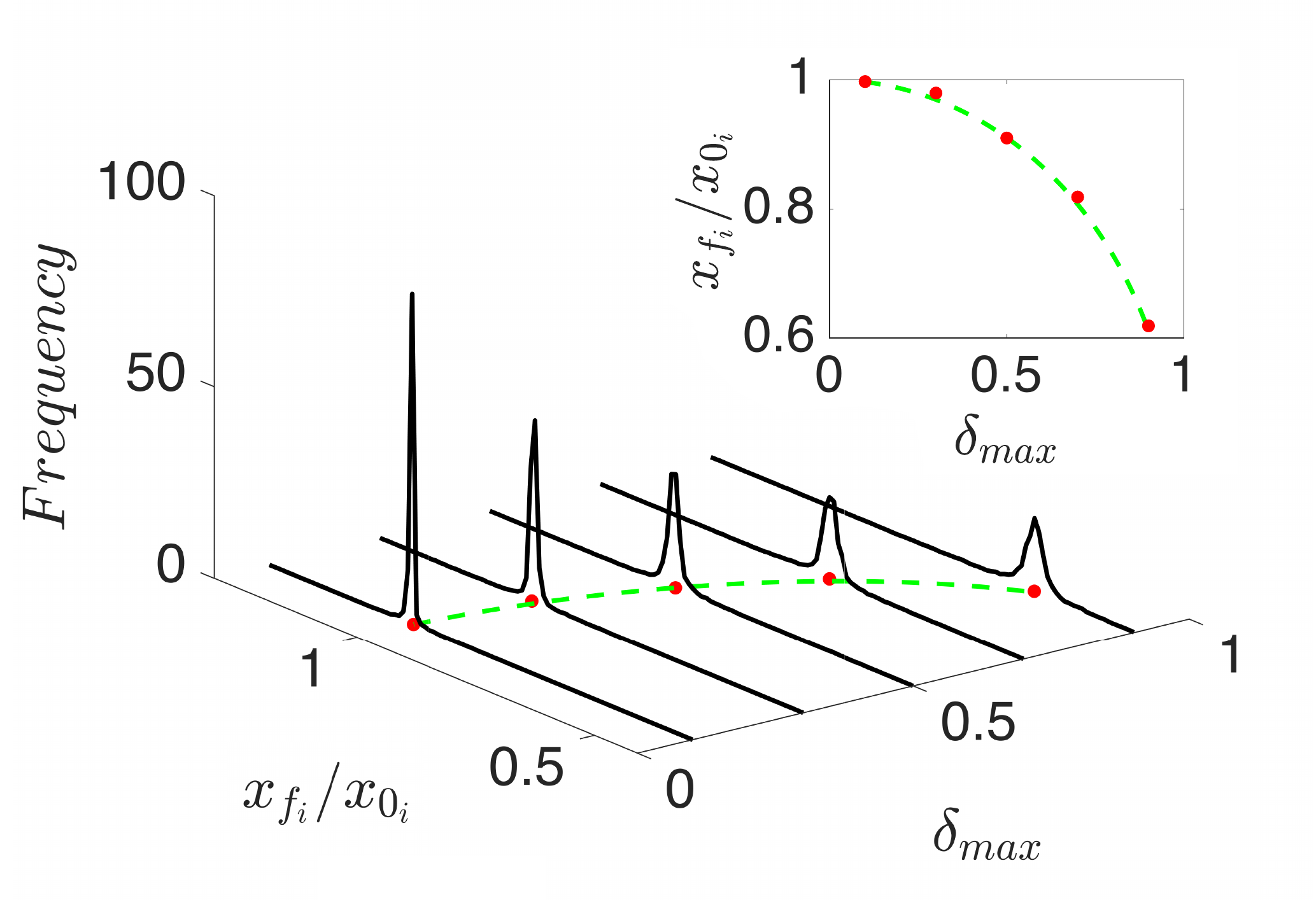}
    \caption{Comparison between theoretical predictions and numerical results. The solid black lines show histograms of  $x_{f_i}/ {x_{0_i}}$ plotted for all nodes of the network. For each of the values of maximum noise $\delta_{max}$ shown, a new network was generated and tested according to the method outlined in the text. Each of the tested networks has $N=2000$ masses and have been created according to the Watts-Strogratz strategy \cite{Watts1998CollectiveNetworks} with rewiring probability $p=0.7$. The dotted green line represents the expected shift per node, $1/\psi$, against $\delta_{max}$. The red dots mark the mean of each of the histograms. A top view of the plot is shown in the inset.}
    \label{fig:1}
\end{figure}

Note that $\psi$ is only a function of the maximum prestrain $\delta_{max}$, and hence the expected shift of each mass, as described by Eq.~\eqref{eq11}, is only affected by $\delta_{max}$. In fact, as it can be seen from the plot in Fig.\ref{fig:1}, numerical simulations of the process described in the previous section match remarkably well with the theoretical predictions of shrinkage even with relatively small networks of size $N=2000$. This also confirms our assumption that the effect of the rigid sphere collisions is well captured by considering the vector $\overrightarrow{B}$ as constant.

It should also be noted from Fig. \ref{fig:1} and Eq.~\eqref{eq10} that $\frac{x_{f_i}}{x_{0_i}}<1$ for the test points shown and therefore the networks are shrinking on average.
The average magnitude of the ENM overall shrinkage can be measured via the radius of gyration  
\begin{equation}
    R_g^2=\frac{1}{N} \sum_{i=1}^N (x_i - x_m)^2
    \label{eq12}
\end{equation}
where $x_m$ refers to the centre of mass of the system. This can used to compute the relative shrinkage $\Delta R_g^2=\frac{(R_{g_f}^2-R_{g_0}^2)}{R_{g_0}^2}$, where $R_{g_0}^2$ and $R_{g_f}^2$ are, respectively, the initial and the final radii of gyration. 
Thanks to Eq.~\eqref{eq11}, the expected value for the final radius of gyration can be written in terms of the shrinkage factor $\psi$ as
\begin{equation}
\langle R_{g_f}^2 \rangle  =\frac{1}{\psi^2} R_{g_0}
\label{eq13}
\end{equation}
with the expected relative shrinkage then reading
\begin{equation}
\langle \Delta Rg^2 \rangle = \frac{1}{\psi^2} - 1
\label{eq14}
\end{equation}

Note that, in the setting considered in this paper, the prestrained elastic networks always shrink on average. In fact, given a maximum prestrain between 0\% and 100\% (i.e.  $\delta_{max} \in (0,1)$), $\psi$ is always greater than one and hence, from Eq.~\eqref{eq14}, $\langle \Delta Rg^2 \rangle<0$.
It should also be remarked that a similar analysis can be performed using the distribution of prestrained spring constants $P(k)$ instead of the distribution of prestrain $P(\delta)$ to obtain the same results. 

\section{Numerical Tests}
 
The analytical results described above were validated against numerical simulations of the exact dynamics Eq.~\eqref{eq1}. Such numerical simulations were carried out also to rigorously assess the effect that other factors, such as network size, topology and the average coordination number, might have on the relative shrinkage $\langle \Delta Rg^2 \rangle$. We define the average coordination number as $\rho_{av}=\frac{L}{2N}$ where $L$ is the total number of links in the network.

\begin{figure}[htb!]
    \centering
   \includegraphics[scale=0.22 ]{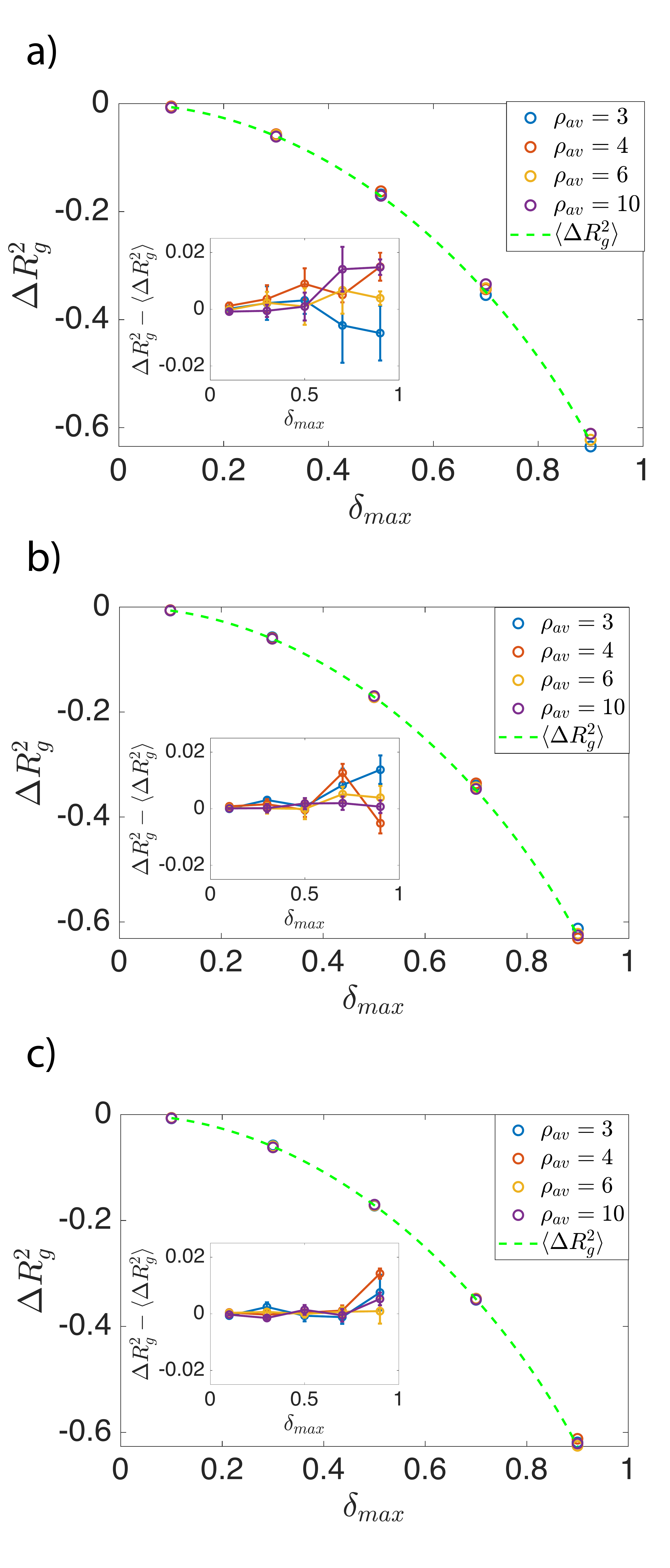}
    \caption{a) Variation of the relative shrinkage $ \Delta Rg^2 <0$ against the maximum prestrain  $\delta_{max}$ for networks of $N=500$ masses while b) and c) show similar results of tests done for $N=2000$ and $N=4000$ respectively. The dotted green line represents the expectation calculated using Eq.~\eqref{eq12} and the solid lines refer to numerical simulations. Networks of different average coordination numbers, $\rho_{av}$ are shown by colour. Each marker (dot) shows the mean of 6 trials with separate realisations of networks, generated according to the Watts-Strogatz (rewiring probability $p=0.7$). The error bars represent the Standard Error. The inset shows the difference between the expected relative shrinkage and the shrinkage from the numerical simulations $\Delta Rg^2 - \langle \Delta Rg^2 \rangle$.}
    \label{fig:2}
\end{figure}

The results of tests performed on random networks, generated with rewiring probability $p=0.7$, of sizes $N=500, 200$ and $4000$ masses are shown in Fig.~\ref{fig:2}. For each of these sizes we tested varying average coordination numbers.
 
As can be seen in Fig.~\ref{fig:2}, neither the size of the network nor the link density have a major effect on the observed $\langle \Delta Rg^2 \rangle$. On the other hand, there is a minor finite size effect that can be appreciated in Fig.~\ref{fig:2} for large values of $\delta_{max}$. Such discrepancy is due to scarcity of samples to accurately estimate average values in Eq.~\eqref{eq8} and vanishes as the number of masses increases. The percentage deviation of the observed shrinkage from the theoretical predictions is about $2\%$ at $\delta_{max}=0.9$ for $N=4000$, while it is about $5\%$ for networks of $N=500$ masses.

Overall, these numerical results confirm the theoretical prediction that the expected shrinkage is only influenced by the magnitude of prestrain $\delta_{max}$.

\subsection{Role of network topology}

\begin{figure}[htb!]
    \centering
    \includegraphics[scale=0.35]{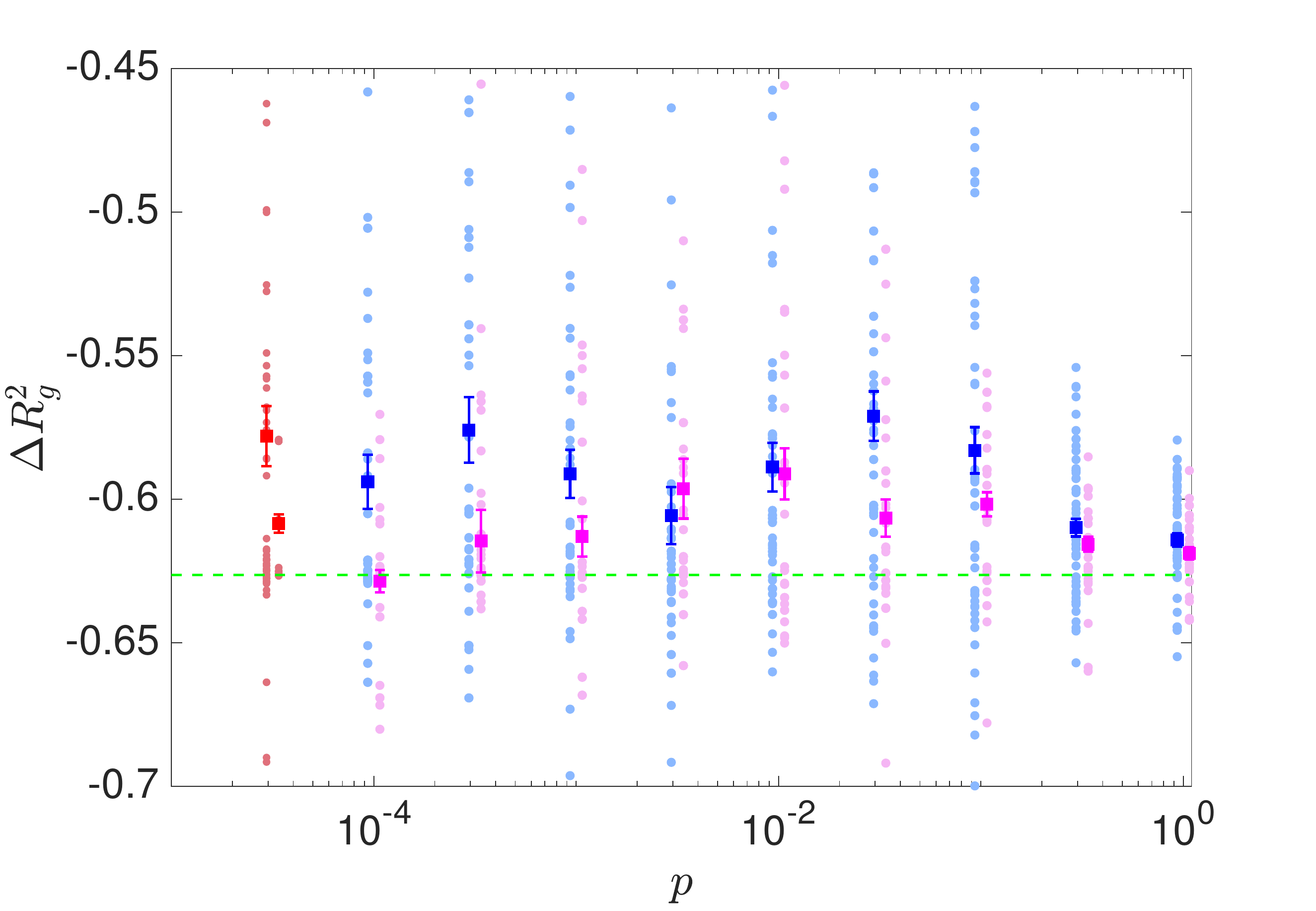}
    \caption{Distribution (mean and standard error) of relative shrinkage $\langle \Delta Rg^2 \rangle$ against the rewiring probability $p$ of the generated Watts-Strogatz networks. 50 trials were tested for each rewiring probability by generating a new network of $N=1000$ (shown in blue) and $N=2000$ (shown in magenta). The prestrain level was set to the maximum $\delta_{max}=0.9$ and the dotted green line is the expectation given by the theoretical analysis. The scatter of the data is shown underneath as dots of a paler colour, to show the overall distribution. The red points refer to tests done with rewiring $p=0$ and have been added on to the same axis for comparison for both $N=1000$ (red left-hand side marker) and $N=2000$ (red right-hand side marker).}
    \label{fig:3}
\end{figure}

In order to investigate the effect of network topology on the expected network shrinkage, several numerical simulations were performed on different classes of networks. Firstly, the effect of small worldliness on the overall network shrinkage was studies via a series of tests  on Watts-Strogatz networks with varying rewiring probability $p$. For each value of $p$, 50 tests were performed and then the overall shrinkage averaged as shown in Fig.~\ref{fig:3}. 
Each network was generated starting from a ring lattice of $N$ nodes with $\rho_{av}=3$ (which translates to average node degree $\langle k \rangle = 6$), and the links then rewired randomly with a probability $p$ while avoiding self loops and duplicated links.
From results displayed in Fig.~\ref{fig:3} it can be seen that the variance of the shrinkage decreases as the rewiring $p$ increases.
The mean values are slightly higher ($7\%$) than the expected value given by the theory for low rewiring $p$ while this difference drops to $2\%$ for large values of $p$. This is compatible with the finite size effect that was observed in Fig.~\ref{fig:2}.

\begin{figure}[htb!]
    \centering
    \includegraphics[scale=0.44]{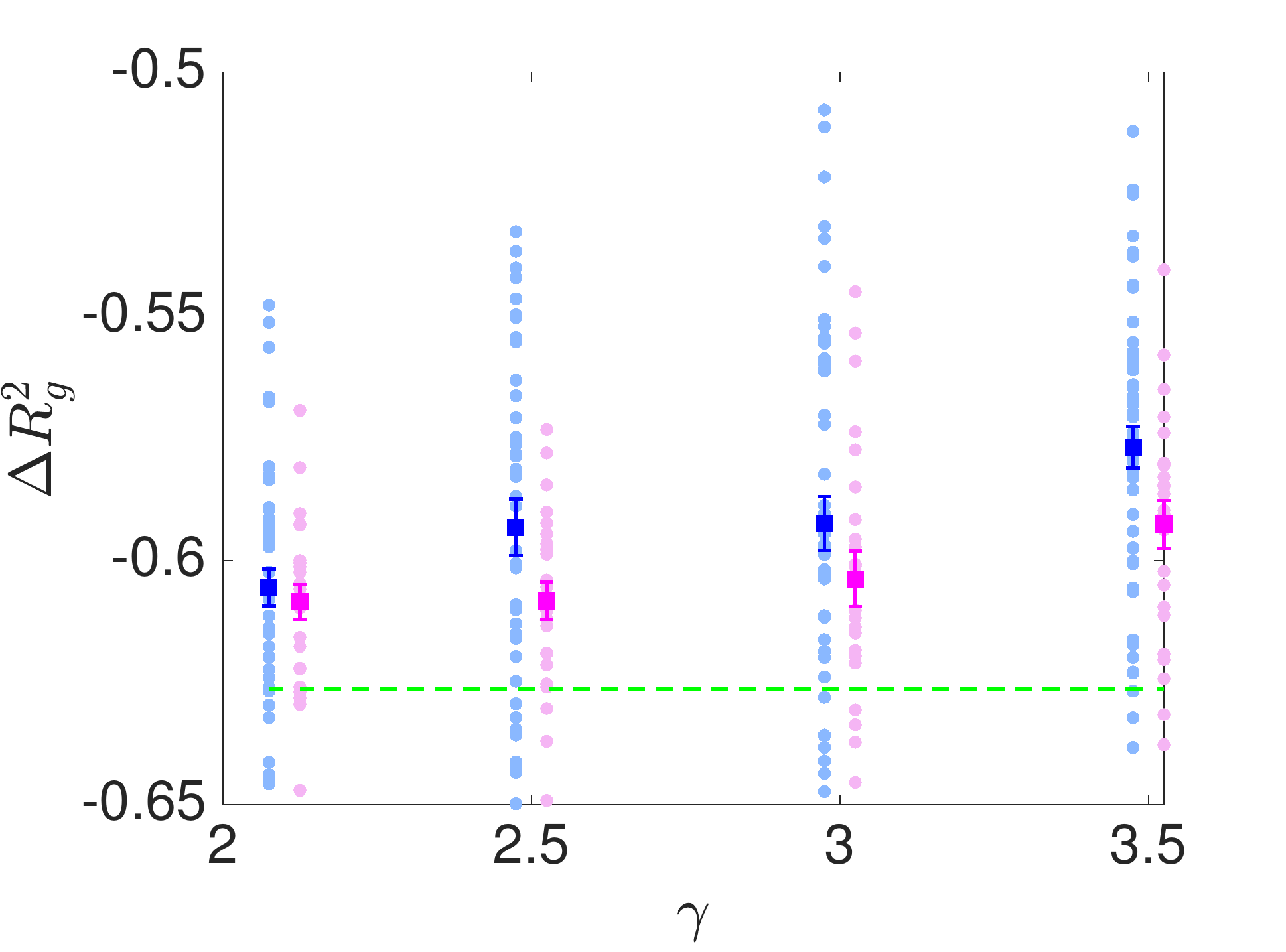}
    \caption{Distribution (mean and standard error) of relative shrinkage $\langle \Delta Rg^2 \rangle$ against the degree exponent $\gamma$ of the generated scale-free networks. 50 trials were tested for each $\gamma$ by generating a new realisation of a network of $N=2000$ for each trail.  The prestrain strength was set to the maximum $\delta_{max}=0.9$ and dotted green line is the expectation given by the theory. $N=2000$ is shown in magenta and thre results of $N=1000$ is shown in blue. The pale dots show the scatter of the data.}
    \label{fig:4}
\end{figure}

Similar tests were carried out for scale-free networks to understand whether structural properties of scale-free networks, such as the degree exponent, have an effect on the observed shrinkage distribution.
To create the test networks, a random degree sequence was generated for $N=2000$ nodes with a given degree exponent $\gamma$.  Vertices or node degree is then assigned to the nodes from this degree sequence, essentially creating stubs or half connections. The configuration model with hidden parameters was then used to connect the stubs avoiding self loops and multiple links in order to create a connected network from the degree sequence \cite{Caldarelli2002ScaleFree}.
Fig.~\ref{fig:4} shows that there is a small tendency for the shrinkage to decrease as the degree exponent $\gamma$ increases. This again could be attributable to the small size of the networks tested not being able to form large enough hubs to see the effect of the exponent or the degree 'cutoff' effect in such networks.

\begin{figure}[htb!]
    \centering
    \includegraphics[scale=0.415]{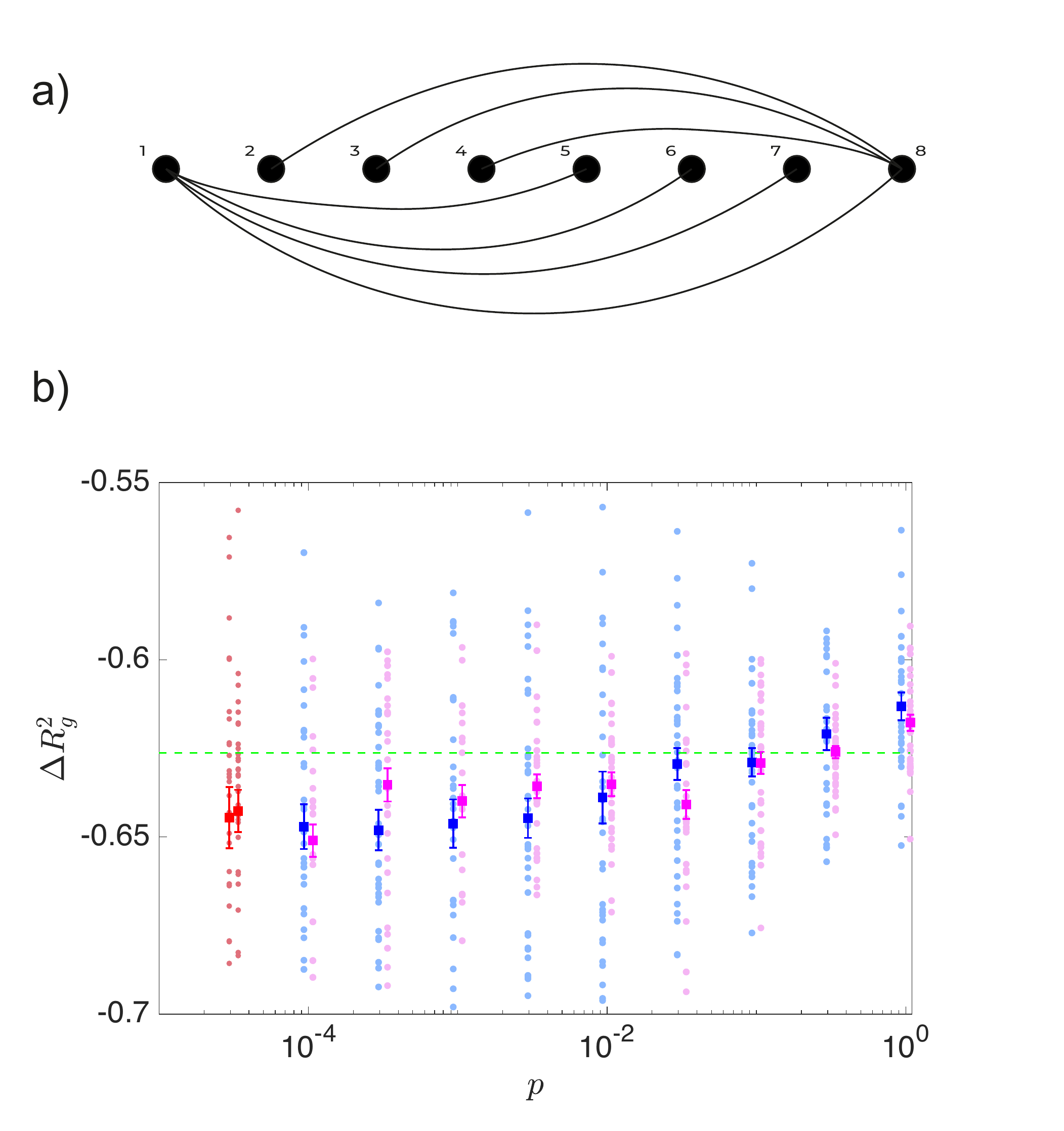}
    \caption{a) Sketch of of an End-Hub network of size $N=8$ with $H=1$ hub on either side. b) Distribution (mean and standard error) of relative shrinkage $\langle \Delta Rg^2 \rangle$ against the rewiring probability of the generated End-Hub networks with $H=3$ hubs per side of the network. 35 trials were tested for each rewiring probability by generating a new network of $N=1000$ nodes (shown in blue) and 16 trials have been tested for $N=2000$ (shown in magenta). The prestrain strength was set to $\delta_{max}=0.9$ and dotted green line is the expectation given by the theory. Pale colour dots  show the scatter of the data.}
    \label{fig:5}
\end{figure}

The last class of networks tested are what we are calling 'End-Hub' networks. These are symmetric networks that have equal number of hubs on either end of the 1D network, and are designed to maximise the link distances. 
For example, if a network of $N=8$ nodes has two hubs with $H=1$ hub on either ends, then the links are such that each hub links to half the nodes of network. This is sketched in Fig.~\ref{fig:5}a for reference.
Further tests were done to assess the effect of rewiring these networks in a fashion similar to that done with the Watts-Strogatz networks. Once again, theoretical predictions match numerical results within $4\%$. 
The comparison between the results shown in Fig. \ref{fig:5}b and those in Fig.~\ref{fig:3} highlights that, while  Watts-Strogatz ENMs have a mean $\langle \Delta Rg^2 \rangle$ that starts above the theoretical predictions and gets closer with increasing randomness, the opposite is true for End-Hub networks. The reason for this is believed to be linked to the initial length of the springs. The regular lattice starts from the shortest possible lengths for any given number of links while End-Hubs starts from the longest links.
As the randomness increases with the rewiring probability we see that both of these trends converge to roughly the same value for $p=1$.

\section{Conclusions}
We proposed and studied a framework for inducing contraction in 1D elastic networks using a random, zero-mean, prestrain. We have analysed the expected behaviour of such systems within the given bounds and  shown that these networks will always contract with the applied prestrain and, furthermore, that for large networks the amount of contraction is only influenced by the magnitude of the prestrain. Through numerical testing we have found the theoretical predictions for the average shrinkage to be robust with networks as small as $N=500$  nodes. However, minor fluctuations were observed around the expected value at high prestrain strengths. This can be attributed to finite size effects, as those fluctuations vanish for larger networks of $N=2000$ or higher. 

In the small $N$ region where fluctuations were significant, we investigated the role played by the topology of the connecting network by numerically testing Watts-Strogatz, scale-free and End-Hub networks. It was found that having order and regularity in the links, as in the case of Watts-Strogatz and End-Hubs with low rewiring probability $p$, can influence the direction of the observed fluctuations. This can be attributed to the average link length, being shorter (Watts-Strogatz) or longer (End-Hubs) than if the network were completely random. For large values of rewiring probability $p$ all topologies converge to similar values of shrinkage.

It should be remarked that, although network topology plays a limited role in the 1D case analysed here, it could have a more prominent role in the behaviour of networks in higher dimensions and represents an interesting avenue for future study.

\section{References}
\bibliographystyle{ieeetr}
\bibliography{references}

%\begin{thebibliography}{99}

%\end{thebibliography}

\end{document}